\documentclass{amsart}
\usepackage{amsmath,amsfonts,amssymb}
\usepackage{mathrsfs,geometry}
\usepackage[T1]{fontenc}
\usepackage[latin1]{inputenc}
\usepackage{url}

\newcommand{\iy}{\ensuremath{\infty}}
\newcommand{\R}{\ensuremath{\mathbf{R}}}

\newcommand{\N}{\ensuremath{\mathbf{N}}}

\newcommand{\E}{\ensuremath{\mathbf{E}}}

\renewcommand{\P}{\ensuremath{\mathbf{P}}}
\newcommand{\e}{\varepsilon}
\newcommand{\st}{\textnormal{ s.t. }}
\renewcommand{\leq}{\leqslant}
\renewcommand{\geq}{\geqslant}

\newcommand{\disp}{\displaystyle}

\newcommand{\mM}{\mathscr{M}}

\newcommand{\mH}{\mathcal{H}}

\newcommand{\norm}[1]{\left\Vert #1\right\Vert}
\newcommand{\ket}[1]{| #1 \rangle}
\DeclareMathOperator{\esssup}{esssup} 

\newtheorem{theo}{Theorem}
\newtheorem{pr}{Proposition}
\newtheorem{lemma}{Lemma}
\newtheorem{rk}{Remark}

\newtheorem{conj}{Conjecture}

\begin{document}
\author{\textsc{Guillaume Aubrun and Ion Nechita}}
\title{Catalytic majorization and $\ell_p$ norms}
\begin{abstract}
An important problem in quantum information theory is the mathematical
characterization of the phenomenon of quantum catalysis: when can the
surrounding entanglement be used to perform transformations of a jointly
held quantum state under LOCC (local operations and classical
communication) ? Mathematically, the question amounts to describe, for a
fixed vector $y$, the set $T(y)$ of vectors $x$ such that 
we have $x \otimes z \prec y \otimes z$ for some $z$, where $\prec$ denotes the
standard majorization relation.

Our main result is that the closure of $T(y)$ in the $\ell_1$ norm can
be fully described by inequalities on the $\ell_p$ norms: $\norm{x}_p
\leq \norm{y}_p$ for all $p \geq 1$. This is a first step towards a
complete description of $T(y)$ itself. It can also be seen as a
$\ell_p$-norm analogue of Ky Fan dominance theorem about unitarily
invariant norms. The proofs exploits links with another quantum
phenomenon: the possibiliy of multiple-copy transformations ($x^{\otimes
n} \prec y^{\otimes n}$ for given $n$). The main new tool is a variant of Cramér's 
theorem on large deviations for sums of i.i.d. random variables.
\end{abstract}

\maketitle

\section{Introduction}

The increasing interest that quantum entanglement has received in the past decade is due, in part, to its use as a \emph{resource} in quantum information processing. We investigate the problem of entanglement transformation: under which conditions can an entangled state $\ket \phi$ be transformed into another entangled state $\ket \psi$ ? We restrict ourselves to LOCC protocols: Alice and Bob share $\ket \phi$ and have at their disposal only \emph{local operations} (such as unitaries $U_A \otimes I_B$ for Alice) and \emph{classical communication}. Nielsen showed in \cite{nielsen_art} that such a transformation is possible if and only if $\lambda_\phi \prec \lambda_\psi$, where ``$\prec$'' is the \emph{majorization} relation and $\lambda_\phi$, $\lambda_\psi$ are the Schmidt coefficients vectors of $\ket \phi$ and $\ket \psi$ respectively. Practically in the same time, Jonathan and Plenio \cite{jp} discovered a striking phenomenon: entanglement can help LOCC communication, without even being consumed. Precisely, they have found states $\ket \phi$ and $\ket \psi$ such that $\ket \phi$ cannot be transformed into $\ket \psi$, but, with the help of a \emph{catalyst} state $\ket \chi$, the transformation $\ket \phi \otimes \ket \chi \rightarrow \ket \psi \otimes \ket \chi$ is possible. When such a catalyst exists, we say that the state $\ket \phi$ is \emph{trumped} by $\ket \psi$ and we write $\lambda_\phi \prec_T \lambda_\psi$. We say then that $\ket \phi$ can be transformed into $\ket \psi$ by entanglement-assisted LOCC or ELOCC. It turns out that the trumping relation is much more complicated that the majorization relation; one can easily check on two given states $\ket \phi$ and $\ket \psi$ whether $\lambda_\phi \prec \lambda_\psi$ is satisfifted or not, but there is no direct way to determine if $\lambda_\phi \prec_T \lambda_\psi$. Later, Bandyopadhyay et al. \cite{bandy} discovered that a similar situation occurs when trying to transform by LOCC \emph{multiple copies} of $\ket \phi$ into $\ket \psi$. It may happen that the transformation $\ket \phi \rightarrow \ket \psi$ is not possible, but when considering $n$ copies, one can transform $\ket \phi^{\otimes n}$ into $\ket \psi^{\otimes n}$. The phenomenon of multiple simultaneous LOCC transformations, or MLOCC, has been intensively studied in the last years and many similarities with ELOCC have been found \cite{dfly,djfy}.


In this note, we make some progress towards a complete characterization of both ELOCC and MLOCC. We show that a set of inequalities involving $\ell_p$ norms (see the remark on Conjecture \ref{conj-nielsen} at the end of the paper) is equivalent to the fact that $\ket \phi$ can be approached by a sequence of states $\ket{\phi_n}$ which are MLOCC/ELOCC-dominated by $\ket \psi$. An important point is that we allow the dimension of $\ket{\phi _n}$ to exceed the dimension of $\ket \phi$. Our proof uses probabilistic tools; we introduce probability measures associated to $\ket \phi$ and $\ket \psi$ and we use large deviation techniques to show the desired result.

Interestingly, the result can be reversed to give a characterization of $\ell_p$ norms that is similar to the Ky Fan characterization  of unitarily invariant norms. We refer the interested reader to Section \ref{sec:kyfan}. The rest of the paper is organized as follows: in Section \ref{sec:notation} we introduce the notation and the general framework of entanglement transformation of bipartite states. We also state our main result, Theorem \ref{th:main}. The theorem is proved in Section \ref{sec:proof}. Conclusions and some directions for further study are sketched in Section \ref{sec:conclusion}. The appendix at the end of the paper contains basic results from large deviation theory needed in the proof of the main theorem.

\smallskip

{\bf Acknowledgement:} we thank the referees for several helpful remarks that improved the presentation of the paper.

\medskip

\section{Notation and statement of the results}\label{sec:notation}
For $d \in \N^*$, let $P_d$ be the set of $d$-dimensional probability vectors : $P_d = \{ x \in \R^d \st x_i \geq 0, \sum x_i=1 \}$. 
If $x \in P_d$, we write $x^{\downarrow}$ for the decreasing rearrangement of $x$, i.e. the vector $x^\downarrow \in P_d$ such that $x$ and $x^\downarrow$ have the same coordinates up to permutation, and $x^\downarrow_i \geq  x^\downarrow_{i+1}$. We shall also write $x_{\max}$ for $x^\downarrow_1$ and $x_{\min}$ for the smallest nonzero coordinate of $x$.

There is an operation on probability vectors that is fundamental in what follows: the tensor product $\otimes$. 
If $x=(x_1,\dots,x_d) \in P_d$ and $x'=(x'_1,\dots,x'_{d'}) \in P_{d'}$, the tensor product $x \otimes x'$ is the vector $(x_i x'_{j})_{ij} \in P_{dd'}$; the way we order the coordinates of $x \otimes x'$ is immaterial for our purposes. We also define the direct sum $x \oplus x'$ as the concatenated vector $(x_1,\dots,x_d,x'_1,\dots,x'_{d'})\in \R^{d+d'}$.  

It $x \in P_d$ satisfies $x_d=0$, it will be useful to identify $x$ with the truncated vector $(x_1,\dots,x_{d-1}) \in P_{d-1}$. This identification induces a canonical inclusion $P_{d-1} \subset P_d$. Thus, every vector $x \in P_d$ can be thought of as a vector of $P_{d'}$ for all $d' \geq d$ by appending $d'-d$ null elements to $x$. We consider thus the set of all probability vectors $P_{<\iy} = \bigcup_{d>0} P_d$. In other words, $P_{<\iy}$ is the set of finitely supported probability vectors.

Let us now introduce the classical majorization relation \cite{mo,bhatia}. If $x,y \in \R^d$ we define the submajorization relation $\prec_w$ as follows
\[x \prec_w y \quad \text{iff.} \quad \forall k \in \{1, \ldots d\}, \, \sum_{i=1}^{k}{x^\downarrow_i} \leq \sum_{i=1}^{k}{y^\downarrow_i},\]
and the majorization relation $\prec$ as
\[ x \prec y \quad \text{iff.} \quad \sum_{i=1}^d x_i = \sum_{i=1}^d y_i \quad \text{and} \quad \forall k \in \{1, \ldots d-1\}, \, \sum_{i=1}^{k}{x^\downarrow_i} \leq \sum_{i=1}^{k}{y^\downarrow_i}. \]

We usually work with probability vectors, for which both relations coincide. However, it will be useful in the proof to work with deficient vectors (of total mass less than 1) and to use submajorization. We write $S_d(y)$ for the set of vectors $x$ in $P_d$ which are majorized by $y$. It is well-known that $S_d(y)$ is a compact convex set whose extreme points are the vectors obtained by permuting the coordinates of $y$; many other characterizations of $S_d(y)$ are known \cite{nielsen_book,daftuar}. This relation behaves well with respect to direct sums and tensor products: $x \prec y$ implies $x \oplus z \prec y \oplus z$ and $x \otimes z \prec y \otimes z$ for any $z \in P_{<\iy}$. The majorization relation has been shown to have a very important role in quantum information. Nielsen has proved \cite{nielsen_art} that a state $\ket \phi$ belonging to Alice and Bob can be transformed into the state $\ket \psi$ by using local operations and classical communication (LOCC) if and only if
\[\lambda_\phi \prec \lambda_\psi,\]
where $\lambda_\phi$ (respectively $\lambda_\psi$) is the vector of eigenvalues of the density matrix for Alice's system when the joint system is in the state $\ket \phi$ (respectively $\ket \psi$). Not long after Nielsen's theorem, Jonathan and Plenio have discovered a very intriguing phenomenon: there exist states $\ket \phi$ and $\ket \psi$ such that the transformation $\ket \phi \rightarrow \ket \psi$ is impossible by LOCC, but, with the aid of a \emph{catalyst state} $\ket \chi$, the transformation $\ket \phi \otimes \ket \chi \rightarrow \ket \psi \otimes \ket \chi$ becomes possible; we say that $\ket \phi$ can be transformed into $\ket \psi$ by \emph{Entanglement-assisted LOCC} or ELOCC. This result has motivated a more complex relation between probability vectors: if $x,y \in P_d$, we say that $y$ trumps $x$ and write $x \prec_T y$ if there exists $z \in P_{<\iy}$ such that $x \otimes z \prec y \otimes z$. It is important to require that the auxiliary vector $z$ (called the catalyst) is finitely supported (see Remark \ref{rk:inf_cat}). Given $y \in P_d$, we write $T_d(y)$ for the set of $d$-dimensional vectors trumped by $y$, that is 
\[ T_d(y) = \{ x \in P_d \st x \prec_T y \} .\]
The set $T_d(y)$ is in general larger than $S_d(y)$ \cite{dk} and much more complicated to describe. Up to now, there is no known simple procedure to decide whether $x \in T_d(y)$ or not. Hence, finding a tractable characterization of the relation $\prec_T$ (or, equivalently, of the set $T_d(y)$) has become an important open problem in quantum information theory \cite{open}. The geometry of $T_d(y)$ has been studied in \cite{daftuar,dk}: it is a bounded convex set that it is neither closed nor open (provided $y$ is not too simple). 
We shall introduce now another important extension of LOCC transformations. Bandyopadhyay et al \cite{bandy} found an example of entangled states $\ket \phi$ and $\ket \psi$ with the property that the LOCC transformation $\ket \phi \rightarrow \ket \psi$ is impossible but, when one tries to transform multiple copies of the states, the transformation $\ket \phi^{\otimes n} \rightarrow \ket \psi^{\otimes n}$ becomes possible. We say that $\ket \psi$ MLOCC-dominates $\ket \phi$. We introduce the analogue of the trumping relation for probability vectors:
\[x \prec_M y \quad \text{iff} \quad \exists n \geq 1 \, \text{s.t.} \, x^{\otimes n} \prec y^{\otimes n},\]
and the set of probability vectors MLOCC-dominated by a given vector $y$:
\[ M_d(y) = \{ x \in P_d \st x \prec_M y \} .\]
Not much is known about the set $M_d(y)$. It has been studied in \cite{dfly} and shown to have many similarities with the set $T_d(y)$: for example it is neither closed nor open in general. One important point is that, for all $y$, we have $M_d(y) \subseteq T_d(y)$ (see \cite{dfly}).

We report progress towards a description of the sets of $M_d(y)$ and $T_d(y)$. The main ingredient of our approach is the following observation. Consider two vectors $x, y \in P_d$. Whether $x \prec y$, $x \prec_M y$, $x \prec_T y$ or not depends only on the non-zero coordinates of $x$ and $y$. Thus, it is possible to $\prec$/$\prec_M$/$\prec_T$-compare vectors of different sizes by appending the necessary amount of zero coordinates to the end of one of them. Hence, it seems more natural (at least from a mathematical point of view) to consider the sets
\[ T_{<\iy}(y) = \{ x \in P_{<\iy}  \st x \prec_T y\} = \{ x \in P_{<\iy} \st \exists z \in P_{<\iy} \st x \otimes z \prec y \otimes z \} = \bigcup_{d' \geq d}T_{d'}(y)\]
and
\[ M_{<\iy}(y) = \{ x \in P_{<\iy} \st x \prec_M y\} = \{ x \in P_{<\iy} \st \exists n \geq 1 \st x^{\otimes n} \prec y^{\otimes n} \} = \bigcup_{d' \geq d}M_{d'}(y).\]
The important point here is that both $T_{<\iy}(y)$ and $M_{<\iy}(y)$ do not depend anymore on the size of $y$, but only on the non-null coordinates of $y$. 
Of course, if $y \in P_d$, $T_d(y)=T_{<\iy}(y) \cap P_d$ and $M_d(y)=M_{<\iy}(y) \cap P_d$; this shows that the sets $T_{<\iy}(y)$ and $M_{<\iy}(y)$ are not closed either in general (otherwise $T_d(y)$ and $M_d(y)$ would be also closed). We then write $\overline{T_{<\iy}(y)}$ and $\overline{M_{<\iy}(y)}$ to denote the closure taken with respect to the $\ell_1$-norm, the natural topology in this setting (see Remark \ref{rk:natural_norm}). Recall that for $p \geq 1$, the $\ell_p$ norm of a vector $x \in P_d$ is defined as
\begin{equation} \label{ell_p_norm} \norm{x}_p = \left( \sum_{i=1}^d x_i^p \right)^{1/p} \end{equation}
and $\norm{x}_{\iy} = \max x_i$. We now come to our main result:

\begin{theo}\label{th:main}
Consider two vectors $x,y \in P_{<\iy}$. The following assertions are equivalent:
\begin{enumerate}
\item[(a)] $x \in \overline{M_{<\iy}(y)}$,
\item[(b)] $x \in \overline{T_{<\iy}(y)}$,
\item[(c)] $\|x\|_p \leq \|y\|_p \, \, \forall p \geq 1$.
\end{enumerate}
\end{theo}
\begin{rk}
Note that instead of demanding that  $\|x\|_p \leq \|y\|_p$ for all $p \geq 1$, it suffices to ask for $x, y \in P_d$ that the inequality holds for all $p \in [1, p_{max}(x,y)]$, where $p_{max}(x,y) = \log d / (\log y_{max} - \log x_{max})$. The inequalities for $p > p_{max}$ follow by simple computation. For such results in a more general setting, see \cite{mitra}.
\end{rk}
\begin{rk}\label{rk:inf_cat}
It is important to see at this point how the set $\overline{T_{<\iy}(y)}$ is related to the set $T_d(y)$. First of all, note that if we drop the closure, we have equality: $T_{<\iy}(y) \cap P_d = T_d(y)$ for $y \in P_d$. However, when taking the $\ell_1$ closure of the left hand side, we obtain a strict inclusion: $\overline{T_d(y)} \subsetneq \overline{T_{<\iy}(y)} \cap P_d$. An example for such a vector is provided by the phenomenon of infinite-dimensional catalysis, discovered by Daftuar \cite{daftuar}. Take $y = (0.5, 0.25, 0.25)$ and $x = (0.4, 0.4, 0.2)$. It is obvious that $x \notin \overline{T_d(y)}$ because $x_3 < y_3$ and the condition $x_d \geq y_d$ is necessary for $x \in \overline{T_d(y)}$. However, there exist an infinite-dimensional catalyst $z = (1-\alpha) (1, \alpha, \alpha^2, \ldots, \alpha^k, \ldots)$, where $\alpha = 2^{-\frac 1 8}$, such that $x \otimes z \prec y \otimes z$ and $||x\otimes z||_p \leq ||y \otimes z||_p$ for all $p \geq 1$. Note that $z$ is $\ell_p$-bounded and thus $\|x\|_p \leq \|y\|_p$ for all $p \geq 1$. By the preceding theorem, we have that $x \in \overline{T_{<\iy}(y)} \cap P_3$. For further remarks on this topic, see Section \ref{sec:conclusion}.
\end{rk}
\begin{rk}\label{rk:natural_norm}
The use of the $\ell_1$ norm is natural in this context from a mathematical point of view since $P_{<\iy}$ is a subset of the norm-closed hyperplane of $\ell_1$ defined by $\sum x_i=1$. Let us explain also how it relates to other physically motivated distances between the approaching states $\ket{\phi_n}$ and the original state $\ket \phi$. Recall that $x$ is the eigenvalue vector of the reduced density matrix corresponding to Alice's (or, equivalently to Bob's) part of the system. From the details of the proof (see also Section \ref{sec:conclusion}), one sees that the size of the approaching vectors $x_n$ increases with $n$. So, in order to compare $\rho$ and $\rho_n$, we have to realize them as density matrices on the same Hilbert space $\mH$. Moreover, we can suppose that the two states are diagonalizable in the same basis (Alice can achieve this by applying a local unitary basis change). As usually, we append the necessary number of zero eigenvalues to $x$ in order to have the same size as $x_n$. We obtain the following equality:
\[\|x-x_n\|_1 = \|\rho - \rho_n\|_{tr}.\]
So, for Alice's part of the system, we obtain a convergence in the trace norm sense. It is well known that the trace norm distance is related to the probability that the two states can be distinguished by some measurement. Moreover, by using some classical inequalities (see \cite{nielsen-chuang}, Chapter 9), the fidelity $F(\rho, \rho_n)$ can be shown to converge to 1.
\end{rk}

\section{A $\ell_p$ version of Ky Fan theorem}\label{sec:kyfan}

In this section, we explain how Theorem \ref{th:main} can be seen as an analogue of Ky Fan dominance theorem. 
We refer to \cite{bhatia} for background. We denote by $\mM_d$ the space of complex $d\times d$ matrices. A norm $|||\cdot|||$ on $\mM_d$ is said to be unitarily invariant if $|||UAV|||=|||A|||$ for all unitary matrices $U,V$. A norm $||\cdot||$ on $\R^d$ is said to be symmetric if
\[ ||(x_1,\dots,x_d)|| = ||(\pm x_{\sigma(1)},\dots,\pm x_{\sigma(d)})|| \]
for all choices of signs in $\{\pm 1\}^d$ and all permutations $\sigma \in \mathfrak{S}_d$.
It is well-known (\cite{bhatia}, Theorem IV.2.1) that unitarily invariant norms on $\mM_d$ are in 1-to-1 correspondance with symmetric norms on $\R^d$ (consider the restriction of $|||\cdot|||$ to diagonal matrices).

Examples of unitarily invariant norms are given by Ky Fan norms, defined for $k=1,2,\dots,d$ by
\[ |||A|||_{(k)} = \sum_{j=1}^k s_j(A) ,\]
where $s_1(A) \geq \cdots \geq s_d(A)$ denote the ordered singular values of a matrix $A$.
The Ky Fan dominance theorem asserts that these norms are extremal among unitarily invariant norms in the following sense: if $A,B$ satisfy $|||A|||_{(k)} \leq |||B|||_{(k)}$ for any $k=1,\cdots,d$, then $|||A||| \leq |||B|||$ for any unitarily invariant norm ; this condition can also be formulated as $s(A) \prec_w s(B)$, where $s(\cdot)$ denotes the vector of singular values of a matrix.

This gives a way to derive an infinite family of inequalities from a finite one. However this may be a too strong requirement and one can wonder what happens for an important special class of unitarily invariant norms: the Schatten $p$-norms (or noncommutative $\ell_p$ norms), defined for $p \geq 1$ by
\[ |||A|||_{p} = \left( \sum_{j=1}^d s_j(A)^p \right)^{1/p}. \]
To state our result, we need to compare matrices of different sizes. If $d<d'$ we identify $\mM_d$ with the top-left corner of $\mM_{d'}$ ; this gives a natural inclusion $\mM_d \subset \mM_{d'}$ and we write $\mM_{< \iy} = \bigcup_{d} \mM_d$. Note that the tensor product of matrices is a well-defined operation on $\mM_{<\iy}$.

\begin{theo} Let $A,B \in \mM_d$. The following are equivalent
\begin{enumerate}
\item $|||A|||_p \leq |||B|||_p$ for all $p \geq 1$.
\item There exists in $\mM_{<\iy}$ a sequence $(A_n)$ so that $\lim_{n \to \iy} |||A_n-A|||_1 = 0$ and $|||A_n^{\otimes n}||| \leq |||B_n^{\otimes n}|||$ for all unitarily invariant norms $|||.|||$ (or, equivalently, so that $s(A_n^{\otimes n}) \prec_w s(B^{\otimes n})$).
\end{enumerate}
\end{theo}

Of course, a main difference between this result and Ky Fan dominance theorem is that condition (ii) here is hard to check and involves infinitely many inequalities. 

\begin{proof}[\it Proof (sketch)]
Because of the bijective correspondance between unitarily invariant norms on matrices and symmetric norms on vectors, it is enough to prove the theorem for positive diagonal matrices. This is almost the content of the equivalence (a)$\iff$(c) of Theorem 1. The only slight remark that we need in order to get condition (2) as stated here is the following: in Lemma \ref{lemma-kuperberg} below, it follows from the proof that we can actually choose the integer $n$ so that $x^{\otimes N} \prec_w y^{\otimes N}$ for any $N \geq n$.
\end{proof}

\section{The proof of the theorem} \label{sec:proof}
We shall prove the sequence of implications (a) $\Rightarrow$ (b) $\Rightarrow$ (c) $\Rightarrow$ (a). The first two are well known; we sketch their proof for completeness. The third is the most difficult one and represents our contribution to the theorem. 

\textbf{(a) $\Rightarrow$ (b)} Because the closure is taken with respect to the same topology ($\ell_1$) for both $\overline{M_{<\iy}(y)}$ and $\overline{T_{<\iy}(y)}$, it is enough to show $M_{<\iy}(y) \subset T_{<\iy}(y)$. Let $x \in M_{<\iy}(y)$ and consider $n$ such that $x^{\otimes n} \prec y^{\otimes n}$. The trick here (see \cite{dfly}) is to use the following $z$ as a catalyst
\[ z = x^{\otimes (n-1)} \oplus x^{\otimes (n-2)} \otimes y \oplus \cdots \oplus x \otimes y^{\otimes (n-2)} \oplus y^{\otimes (n-1)} . \]
For simplicity we do not normalize $z$, but this is irrelevant. The vector $z$ has been constructed such that 
\[ x \otimes z = x^{\otimes n} \oplus w \textnormal{ and } y \otimes z = y^{\otimes n} \oplus w, \]
where $w$ is the same in both expressions. This implies that
$x \otimes z \prec y \otimes z$, i.e. $x \in T_{<\iy}(y)$.

\textbf{(b) $\Rightarrow$ (c)} Let $z \in P_{<\iy}$ be the catalyst for $x \prec_T y$: $x \otimes z \prec y \otimes z$. A function $\varphi : \R^d \to \R$ is said to be Schur-convex if $a \prec b$ implies $\varphi(a) \leq \varphi(b)$. It is well-known (see \cite{mo,nielsen_book}) that if $h : \R \to \R$ is a convex function, then $\varphi : x \mapsto \sum_{i=1}^d h(x_i)$ is Schur-convex.
Consequently, the functions $x \mapsto \|x\|_p^p$ are Schur-convex for $p \geq 1$. Moreover, they satisfy the identity $\|x \otimes z\|_p=\|x\|_p \|z\|_p$, and similarly for $y$. Since $\|z\|_p$ is finite, we get that $\|x\|_p \leq \|y\|_p$. To show that the same is true for $x \in \overline{T_{<\iy}(y)}$, it suffices to check that the set of $x \in \ell_1$ such that $ \|x\|_p \leq \|y\|_p$ is norm-closed; this follows from the inequality $\|\cdot\|_p \leq \|\cdot\|_1$.

\textbf{(c) $\Rightarrow$ (a)} We will adapt some techniques used by G. Kuperberg in a slightly different context \cite{kuperberg}. In our proof, we allow deficient vectors, i.e. vectors with total mass smaller than 1, and we use submajorization.

As in \cite{kuperberg}, we associate to a positive vector $x \in \R^d$ the measure $\mu_x = \sum_{i=1}^{d}{x_i \delta_{\log x_i}}$, where $\delta_z$ is the Dirac measure at point $z$. The basic property is that the tensor product operation of vectors corresponds to the convolution of associated measures:
\[\mu_{x \otimes y} = \mu_x * \mu_y.\]
The convolution of two measures $\mu$ and $\nu$ is defined by the relation
\[\mu * \nu(A) = (\mu \times \nu)\left(\{(x,y) \in \R^2 : x+y \in A\}\right).\]
Moreover, if $\mu$ and $\nu$ are probability measures and $X_\mu$ and $X_\nu$ denote independent random variables with laws respectively $\mu$ and $\nu$, then $\mu*\nu$ is the law of $X_\mu+X_\nu$.

The following lemma gives a way to prove majorization using comparison of the tails of the associated measures

\begin{lemma}
\label{lemma-vector-measure}
Let $x$ and $y$ be two vectors of $\R^d$ with non-negative components. Consider the measures $\mu_x$ and $\mu_y$ associated with $x$ and $y$. Assume that, for all $t \in \R$, $\mu_x[t, \infty) \leq \mu_y[t, \infty)$. Then $x \prec_w y$.
\end{lemma}

\begin{proof}
Note that
\[\mu_x[t, \infty) = \sum_{i:\log x_i \geq t}{x_i} = \sum_{i:x_i \geq \exp(t)}{x_i}.\]
Thus, for all $u > 0$, $\sum_{i:x_i \geq u}{x_i} \leq \sum_{i:y_i \geq u}{y_i}$. 
For simplicity, we assume first that all coordinates of $y$ are distinct.
We will show by induction on $k \in \{1, \ldots, d\}$ that $\sum_{i=1}^{k}{x^\downarrow_i} \leq \sum_{i=1}^{k}{y^\downarrow_i}$. For the first step; use $u=y^\downarrow_1$ to conclude that $x^\downarrow_1 \leq y^\downarrow_1$.
Now, fix $k \in \{1, \ldots, d-1\}$ and suppose that $\sum_{i=1}^{k}{x^\downarrow_i} \leq \sum_{i=1}^{k}{y^\downarrow_i}$. If $x^\downarrow_{k+1} \leq y^\downarrow_{k+1}$, the induction step is obvious. If 
$x^\downarrow_{k+1} > y^\downarrow_{k+1}$, we use $u=x^\downarrow_{k+1}$ to get
\[ \sum_{i=1}^{k+1} {x^\downarrow_i} \leq \sum_{i:x_i \geq x^\downarrow_{k+1}} x_i \leq \sum_{i:y_i \geq x^\downarrow_{k+1}} y_i \leq \sum_{i:y_i \geq y^\downarrow_{k+1}} y_i = \sum_{i=1}^{k+1} {y^\downarrow_i} .\]
This completes the induction when $y$ has distinct coordinates. The general case follows by approximating $y$ by $y+\e_n$, where $(\e_n)$ is a suitable sequence of positive vectors tending to $0$. The approximation is possible since the set of vectors $y$ majorizing a fixed $x$ is closed.
\end{proof}

We now get to the key lemma in our argument. We shall use a slightly modified version of Cramér large deviations theorem --- see Appendix. 

\begin{lemma}
\label{lemma-kuperberg}
Let $x,y$ in $\R^d$, with nonnegative coordinates. Assume that for any $1 \leq p \leq \iy$, we have the {\bf strict} inequality $\norm{x}_p < \norm{y}_p$. Then there exists an integer $n$ such that $x^{\otimes n} \prec_w y^{\otimes n}$.
\end{lemma}

\begin{proof}
Consider $x$ and $y$ satisfying the hypotheses of the lemma. We can assume by multiplying both vectors by a positive constant $K$ that $\|y\|_1=1$. Let $p = 1-\|x\|_1>0$. We introduce the measures $\mu_x$ and $\mu_y$ associated to $x$ and $y$; $\mu_y$ is a probability measure but $\mu_x$ is not, so we add  a mass at $-\iy$ by setting $\overline{\mu}_x = \mu_x + p \delta_{-\iy}$. Let $X$ and $Y$ be random variables distributed according to $\overline{\mu}_x$ and $\mu_y$ respectively. We denote by $(X_n)$ (resp. $(Y_n)$) a sequence of i.i.d. copies of $X$ (resp. Y). 
We are going to show that for $n$ large enough
\begin{equation}
\label{conclusion-proba}
 \forall t \in \R, \, \P ( X_1 + \dots + X_n \geq nt) \leq \P( Y_1 + \dots + Y_n \geq nt) .
\end{equation} 
 This is equivalent to showing that 
\[
\int_{nt}^{\infty}{d\mu_x^{*n}} = \int_{nt}^{\infty}{d\overline{\mu}_x^{*n}} \leq \int_{nt}^{\infty}{d\mu_y^{*n}},
\]
which, by the previous lemma implies $x^{\otimes n} \prec_w y^{\otimes n}$.
Note that the asymptotic behavior of the quantities appearing in \eqref{conclusion-proba} is governed by Cramér's theorem. Let $f_n(t) = \P ( X_1 + \dots + X_n \geq nt)^{1/n}$ and $g_n(t) = \P( Y_1 + \dots + Y_n \geq nt)^{1/n}$. Applying Cramér's theorem (see Appendix), we obtain
\[ f(t) := \lim_{n \to \iy} f_n(t)  = \begin{cases} 1-p \textnormal{ if } t \leq \E(X | X \neq -\iy) \\ e^{-\Lambda_X^*(t)} \textnormal{ otherwise. } \end{cases} \]

\[ g(t) := \lim_{n \to \iy} g_n(t)  = \begin{cases} 1 \textnormal{ if } t \leq \E(Y) \\ e^{-\Lambda_Y^*(t)} \textnormal{ otherwise. } \end{cases} \]

Note also that the log-Laplace of $X$, defined for $\lambda \in \R$ by $\Lambda_X(\lambda) = \log 
\E e^{\lambda X}$, is related to the $\ell_p$ norms of $x$:
\[ \forall \lambda \geq 0, \ \ \ \Lambda_X(\lambda) = \log \|x\|_{\lambda+1}^{\lambda+1} .\]
The same holds for $Y$: $\Lambda_Y(\lambda) = \log \|y\|_{\lambda+1}^{\lambda+1}$ and thus we have $\Lambda_X(\lambda) < \Lambda_Y(\lambda)$ for $\lambda \geq 0$.

Let $M_X=\esssup X = \log \|x\|_{\iy}$ and $M_Y=\esssup Y = \log \|y\|_{\iy}$ ; by hypothesis $M_X<M_Y$. First of all, note that $f_n(t) = 0$ for $t \geq M_X$, so it suffices to show that $f_n \leq g_n$ on $(-\iy, M_X]$, for $n$ large enough.
We claim that $f<g$ on $(-\iy, M_Y)$, and thus on $(-\iy,M_X]$. Indeed, for $\E(Y) \leq t < M_Y$, the supremum in the definition of $\Lambda_Y^*(t)$ is attained at a point $\lambda_0 \geq 0$ (cf Appendix), so we have that
\[f(t) \leq e^{\disp -(\lambda_0t - \Lambda_X(\lambda_0)) } < e^{\disp -(\lambda_0t - \Lambda_Y(\lambda_0))} = g(t),\]
where the \textbf{strict} inequality follows from the fact that $\Lambda_X(\lambda) < \Lambda_Y(\lambda)$, for all $\lambda \geq 0$. For $t < \E(Y)$, $g(t) = 1$ and $f(t) \leq 1-p < 1$. Moreover, the functions $f$ and $g$ admit finite limits in $-\iy$: $\lim_{t \rightarrow -\iy} f(t) = 1-p$ and $\lim_{t \rightarrow -\iy} g(t) = 1$. Thus, on the compact set $[-\iy, M_X]$, the functions $f$ and $g$ are well-defined, non-increasing, continuous and satisfy $f < g$. 

We now use the following elementary fact: if a sequence of non-increasing functions defined on a compact interval $I$ converges pointwise towards a continuous limit, then the convergence is actually uniform on $I$ (for a proof see \cite{ps} Part 2, Problem 127; this statement is attributed to P\'olya or to Dini depending on authors). We apply this result to $(f_n)$ and $(g_n)$ on the interval $I= [-\iy, M_X]$ to conclude that the convergence is uniform for both sequences.
As $f < g$, we can therefore find $n$ large enough such that $f_n \leq g_n$ on $I$, and thus on $\R$. This is equivalent to \eqref{conclusion-proba} and completes the proof of the lemma.
\end{proof}

\begin{rk} 
It is possible to avoid the use of Cramér's theorem by using low-technology estimates on large deviations probability instead, as done in \cite{kuperberg}. This requires additional care to get the required uniform bounds and slightly obfuscates the argument. The only advantage is to give explicit bounds for the value of $n$ in Lemma \ref{lemma-kuperberg}, which our compactness argument does not. These bounds are quite bad anyway, and for example do not allow to replace the $\ell_1$-closure in the main theorem by a $\ell_p$-closure for some $p<1$.

\end{rk}

\textbf{Proof of (c) $\Rightarrow$ (a)} (continued)
Recall that $x$ and $y$ are such that $\|x\|_p \leq \|y\|_p$ for any $p \geq 1$ and that we want to find, for any $\e >0$ small enough, a vector $x_\e \in M_{<\iy}(y)$ such that $\|x-x_\e\|_1 \leq \e$. Let $d_x$ (resp. $d_y$) be the number of nonzero coordinates of $x$ (resp. $y$). We proceed as follows : let $0< \e < 2d_x x_{min}$ and consider the (deficient) vector $x'_\e$ obtained from $x$ by subtracting $\e/2d_x$ to each of its nonzero coordinates. This implies that $x'_\e$ is a positive vector, $\|x-x'_\e\|_1 = \e/2$ and that $x'_\e$ satisfies the hypotheses of Lemma \ref{lemma-kuperberg}. Applying the lemma, we obtain the existence of an integer $n$ such that $(x'_{\e})^{\otimes n} \prec_w y^{\otimes n}$. 

Remember that $x'_\e$ is deficient; we now enlarge it into a vector $x_\e \in P_{<\iy}$ by adding mass $\e/2$. But since we want to keep the property $x_{\e}^{\otimes n} \prec_w y^{\otimes n}$ (which is identical to $x_{\e}^{\otimes n} \prec_w y^{\otimes n}$), a safe way to do this is to add a large number of coordinates, each of them being very small. More precisely, let $x_\e = x'_\e \oplus \delta^{\oplus D}$, where $\delta D = \e/2$ and
$\delta$ is a positive number such that $\delta (x'_\e)_{\max}^{n-1} \leq \min( (x'_\e)_{\min}^n, y_{\min}^n)$. We claim that $x_{\e}^{\otimes n} \prec y^{\otimes n}$, that is, for any $k \geq 1$,
\begin{equation} \label{check_def_maj} \sum_{i=1}^k (x_\e^{\otimes n})^{\downarrow}_i \leq \sum_{i=1}^k (y^{\otimes n})^{\downarrow}_i .
\end{equation}
Indeed, $\delta$ has been chosen so that the $d_x^n$ largest coordinates of $x_\e^{\otimes n}$ are exactly the coordinates of 
$(x'_{\e})^{\otimes n}$, so when $1 \leq k \leq d_x^n$, \eqref{check_def_maj} follows from the relation $(x'_{\e})^{\otimes n} \prec_w y^{\otimes n}$. If $d_x<k\leq d_y^n$, the inequality also holds since the choice of $\delta$ guarantees $(x_\e^{\otimes n})^{\downarrow}_k \leq (y^{\otimes n})^{\downarrow}_k$. Finally if $k \geq d_y^n$, \eqref{check_def_maj} holds trivially since the right-hand side equals 1.

In conclusion, $x_{\e}^{\otimes n} \prec y^{\otimes n}$, and thus $x_\e \in M_{<\iy}(y)$. But $x_\e$ has been constructed such that $\|x-x_\e\|_1 \leq \e$ and thus $x \in \overline{M_{<\iy}(y)}$ which completes the proof of the theorem.

\section{Conclusion and further remarks}\label{sec:conclusion}
In conclusion, we are able to give a nice description of the $\ell_1$-closure of the set $T_{<\iy}(y)$. However, this closure may be substantially larger than the usual closure $\overline{T_d(y)}$ in $P_d$, and requires approximation by vectors with growing support. Our result can be seen as a contribution to a conjecture attributed to Nielsen \cite{daftuar}:
\begin{conj}
\label{conj-nielsen}
Fix a vector $y \in P_d$. Then a vector $x \in P_d$ belongs to $\overline{T_d(y)}$ if and only if the following conditions are verified.
\begin{enumerate}
\item[(1)] For $p \geq 1$, $\|x\|_p \leq \|y\|_p$.
\item[(2)] For $0 < p \leq 1$, $\|x\|_p \geq \|y\|_p$.
\item[(3)] For $p < 0$, $\|x\|_p \geq \|y\|_p$.
\end{enumerate}
\end{conj}

M. Klimesh announced a proof of this conjecture in a short communication \cite{klimesh_conf}, but the solution has not appeared in print yet. However, his methods are different from our approach (private communication). Note that the definition of $\|\cdot\|_p$ given in \eqref{ell_p_norm} is extended to any $p \in \R^*$. For $p<1$, $\|\cdot\|_p$ is not a norm in the usual sense.
We have shown that the condition (1) above is equivalent to $x \in \overline{T_{<\iy}(y)}$. Notice however that $\overline{T_{<\iy}(y)}$ is in general larger than $\overline{T_d(y)}$; note also that the set of $x \in P_d$ that satisfy conditions (1--3) is closed.
The ``only if'' part of the conjecture follows from standard convexity/concavity properties of functionals $\| \cdot \|_p$, see \cite{nielsen_book,daftuar}.

This question also appears in \cite{dfly} where it is formulated using the Rényi entropies. For any real $p \neq 1$, the $p$-Rényi entropy is defined for $x \in P_d$ as 
\[ H_p(x) = \frac{\textnormal{sgn}(p)}{p-1} \log_2 \left( \sum_{i=1}^d x_i^p \right). \]
The limit case $p=1$ corresponds to the usual entropy. The conditions (1--3) of the conjecture can be concisely reformulated as ``$H_p(x) \leq H_p(y)$ for all $p$''.

An intermediate notion is the following: for $y \in P_d$, let $\overline{T_{<\iy}(y)}^b$ be the set of vectors $x \in P_d$ such that there is a sequence $(x_n)$ in $T_{<\iy}(y)$ tending to $x$, with a \textbf{uniform} bound on the size of the support of $x_n$. We think that a description of $\overline{T_{<\iy}(y)}^b$ could be related to the set of vectors which satisfy conditions (1) and (2) --- but not necessarily (3) --- in Conjecture \ref{conj-nielsen}.
 

There is one more consequence of our main theorem we would like to discuss. Recall that when defining catalysis, we insisted on the fact that the catalyst should be finitely-supported. Let $P_{\iy} \subset \ell_1$ be the set of infinite-dimensional probability vectors, and for $y$ in $P_{<\iy}$, define the set $T'(y)$ of (finitely supported) vectors trumped by $y$ using infinite catalysts:
\[ T'(y) = \{ x \in P_{<\iy} \st \exists z \in P_{\iy} \st x \otimes z \prec y \otimes z \}. \]
As shown in \cite{daftuar} (Section 4.3), in general $T_{<\iy}(y) \neq T'(y)$. However, since $x \in T'(y)$ implies $\|x\|_p \leq \|y\|_p$ for all $p \geq 1$, it follows from our main theorem that $\overline{T_{<\iy}(y)} = \overline{T'(y)}$.

\section{Appendix: On Cramér's theorem}\label{appendix_cramer}

We review here some facts from large deviations theory. A complete reference for all the material contained here is \cite{dz}.
Let $X$ be a random variable taking values in $[-\iy,\iy)$. We allow $X$ to equal $-\iy$ with positive probability; this is a nonstandard hypothesis. We however exclude the trivial case $\P(X=-\iy)=1$. We write $\E$ for the expectation. We assume also that the conditional expectation $\E(X | X \neq -\iy)$ is finite.
The cumulant generating function $\Lambda_X$ of the random variable $X$ is defined for any $\lambda \in \R$ by
\[ \Lambda_X(\lambda) = \log \E e^{\lambda X} .\]
It is a convex function taking values in $(-\iy,+\iy]$. Its convex conjugate $\Lambda_X^*$, sometimes called the Cramér transform,  is defined as
\begin{equation} \label{cramertransform} \Lambda_X^*(x) = \sup_{\lambda \in \R} \lambda x - \Lambda_X(\lambda). \end{equation}
Note that $\Lambda_X$ is a smooth and strictly convex function on $[0,+\iy]$. Moreover, $\Lambda_X'(0) = \E(X | X \neq -\iy)$ and $\lim_{\lambda \to +\iy} \Lambda_X'(\lambda) = \textnormal{esssup}(X)$. Consequently, for any $x$ such that $\E(X | X \neq -\iy) < x < \textnormal{esssup}(X)$, the supremum in \eqref{cramertransform} is attained at a unique point $\lambda \geq 0$. We now state Cramér's theorem in a suitable formulation

\begin{pr}
\label{cramer}
Let $X$ be a $[-\iy,+\iy)$-valued random variable such that $\Lambda_X(\lambda) < +\iy$ for any $\lambda \geq 0$. Let $(X_i)$ be a sequence of i.i.d. copies of $X$. Then for any $t \in \R$
\[ \lim_{n \to \iy} \frac{1}{n} \log \P ( X_1 +\dots +X_n \geq tn) = \begin{cases} \log \P( X \neq - \iy) \textnormal{ if } t \leq \E(X | X \neq -\iy) \\ -\Lambda_X^*(t) \textnormal{ otherwise. } \end{cases} \]
\end{pr}

\begin{proof}
Let $\hat{X}$ denote the random variable $X$ conditioned to be finite, that is for any Borel set $B \subset \R$
\[ \P ( \hat{X} \in B) = \frac{1}{1-p} \P( X \in B) ,\]
where $p = \P(X = -\iy)$. A consequence of the classical Cramér theorem (\cite{dz}, Corollary 2.2.19) states that
\begin{equation}\label{eq:cramer}
	\forall t \in \R, \quad \lim_{n \to \iy} \frac 1 n \log \P(\hat{X}_1 + \dots + \hat{X}_n \geq tn) = - \inf_{s \geq t} \Lambda_{\hat X}^*(s).
\end{equation}
One checks that $\Lambda_{\hat{X}} = \Lambda_X - \log (1-p)$, and consequently 
\begin{equation}\label{eq:Lambdas}
	\Lambda_{\hat{X}}^* = \Lambda_X^* + \log(1-p). 
\end{equation}
Note also that
\begin{equation}\label{eq:Probs}
	\P (X_1 +\dots+X_n \geq tn) = (1-p)^n \P( \hat{X}_1 + \dots + \hat{X}_n \geq tn).
\end{equation}

Finally, note that the infimum on the right hand side of (\ref{eq:cramer}) is null for $t \leq \E(\hat X)$ and equals $\Lambda_{\hat X}^*(t)$ for $t > \E(\hat X)$. This follows from the fact that the convex function $t \mapsto \Lambda^*_{\hat{X}}(t)$ attains its zero minimum at $t = \E(\hat X)$ and is increasing for $t \geq \E(\hat X)$. Thus, we can rewrite equation (\ref{eq:cramer}) as:
\begin{equation}\label{eq:Cramer2}
	\lim_{n \to \iy} \frac{1}{n} \log \P ( \hat{X}_1 +\dots +\hat{X}_n \geq tn) = \begin{cases} 0 \textnormal{ if } t \leq \E(\hat{X}) \\ -\Lambda_{\hat{X}}^*(t) \textnormal{ otherwise. } \end{cases}
\end{equation}
The proposition follows from the equations (\ref{eq:Lambdas}), (\ref{eq:Probs}) and (\ref{eq:Cramer2}).
\end{proof}

\bigskip

Address : \\
Université de Lyon, \\
Université Lyon 1, \\
CNRS, UMR 5208 Institut Camille Jordan, \\
Batiment du Doyen Jean Braconnier, \\
43, boulevard du 11 novembre 1918, \\
F - 69622 Villeurbanne Cedex, \\
France\\
\\
Email: aubrun@math.univ-lyon1.fr, nechita@math.univ-lyon1.fr

\end{document}